\documentstyle[preprint,prl,aps]{revtex}
\begin{document}
\draft
\preprint{}
\title{Survival Probability of a Gaussian Non-Markovian
Process: Application to the T=0 Dynamics
of the Ising Model}

\author{Satya N. Majumdar$^1$ and Cl\'ement Sire$^2$}
\address{$^1$Yale University,
Department of Physics,  Sloane Physics Laboratory\\
New Haven, CT-06520-8120, USA (e-mail: satya@cmphys.eng.yale.edu)\\
$^2$Laboratoire de Physique Quantique (UMR 5626 du CNRS),
Universit\'e Paul Sabatier\\
31062 Toulouse Cedex, France (e-mail: clement@irsamc2.ups-tlse.fr)}
\maketitle
\begin{abstract}
We study the decay of the probability for a \it non-Markovian \rm stationary
Gaussian walker not to cross the origin up to time $t$.
This result is then used to evaluate the fraction of spins
that do not flip up to time $t$ in the zero temperature Monte-Carlo spin
flip dynamics of the Ising model. Our results are compared to
extensive numerical simulations.

\end{abstract}

\newpage

Gaussian processes are amongst the most widely studied stochastic processes
in various branches of science \cite{vK}. However, there are still simple but
important questions associated with a Gaussian process that are nontrivial
to compute. One such question is the following: consider a stationary
Gaussian process $X(\tau)$ with zero mean and a prescribed correlator. What
is the probability $P(\beta)$ that $X(\tau)$ does not cross the origin
$X=0$ between $\tau=0$ and $\tau=\beta$? This quantity, although simple
and natural, turns out to  be quite nontrivial to compute\cite{CH}.
Why is this so? A little thought
shows that this quantity probes high order correlations in time of the
dynamics and it depends on the whole history of time evolution of the
system. In this Letter, we would like to address this
question. It turns out that the solution to this problem has very wide
applications in various other problems in Physics. For example, a related
question arises naturally in the context of zero temperature
Monte-Carlo dynamics in any spin system: what is the probability that a spin
does not flip up to time $t$ \cite{BD1,D1,sta,BD2,AB1,MH}.
Similar questions also arise in the study of the fraction
of lattice sites that remain unvisited up to time
$t$ by a random walker or by chemical species in
a generic reaction diffusion system \cite{JC}.

In this Letter we restrict ourselves to one such application, namely
the simple case of the Ising model.
In the zero temperature dynamics of the Ising model, domains of opposite
spins grow with time. At late times, the system is characterized by a
single length scale (typical size of a growing domain) $L(t)\sim t^{1/2}$
\cite {AB}. The fraction $P(t)$ of spins that remain unflipped up to time
$t$ decays as $P(t)\sim L(t)^{-\theta}$ for large
time $t$\cite{BD1}, where $\theta$ is a universal,
dimension dependent, non-equilibrium exponent.
Analytical computation of $\theta$ seems to be extremely nontrivial. Even in
$d=1$, where the Glauber dynamics is exactly solvable, the calculation of
$\theta$ turned out to be a real tour de force, achieved recently by Derrida
$et$ $al.$ \cite{BD1}. They found $\theta_{1d}=3/4$. However, their
technique is special to $d=1$ and seems impossible to extend to higher
dimensions, where only numerical estimates of $\theta$ are available
\cite{D1,sta,BD2}. Therefore it is highly desirable to obtain an approximate
analytical method to determine $\theta$, that we now present.

Let us start with a stationary Gaussian process $X(\tau)$, with zero mean
and a correlator $\langle
X({\tau}_1)X({\tau}_2)\rangle= f(\tau_1 -\tau_2)$. The probability $P(\beta)$
of not crossing the origin up to $\tau=\beta$ is expected to decay as $\exp
(-{\bar \theta}\beta)$ for large $\beta$ (at least when $f(\tau)$ decays
exponentially for large $\tau$). We would like to calculate ${\bar
\theta}$ since, as we will see later, ${\bar \theta}$ is related to the
exponent $\theta$ of the Ising model. If the Gaussian process is Markovian,
for which $f$ is necessarily of the form $f(\tau) = \exp(-\bar\lambda|
\tau|)/2 \bar \lambda$ \cite{SC},
it is possible to show by various methods \cite{Fell,SC,SM1} that ${\bar
\theta}=\bar\lambda $ exactly (see below also). For any other form of $f$, the
process is non-Markovian (i.e., history dependent) and ${\bar \theta}$ is hard
to compute. In fact, ${\bar \theta}$ depends very sensitively on the full
function $f(\tau)$ and not just on its asymptotic properties. Keeping the
Ising problem in mind, we will restrict ourselves only to the class of
non-Markovian processes for which (i) $f'(0^\pm)\neq 0$  and (ii) $f(\tau)\sim
\exp (-{\bar \lambda} \tau)$ for large $\tau$. For convenience, we will then
normalize $f$, setting $f'(0^\pm)=\mp 1/2$ after a proper rescaling of $X$,
such that its Fourier transform satisfies $\omega^2f(\omega)\to 1$, for large
$\omega$.

To illustrate the explicit history dependent nature of the non-Markovian
process, it is useful to write its associated Langevin equation:
\begin{equation}
\frac{dX}{d\tau}=-{\bar \lambda} X +\eta+\int_{-\infty}^\tau J(\tau-\tau')
\eta(\tau')d\tau'
\label{one}
\end{equation}
where $\eta(\tau)$ is a Gaussian white noise with $\langle \eta(\tau)
\eta({\tau}') \rangle=\delta(\tau-\tau')$, and $J$ is a causal ($J(\tau)=0$,
for $\tau<0$) and integrable function. The history dependence is explicitly
encoded in the kernel $J$. For $J=0$, Eq. (1) describes a Markov
process with $f(\tau) =\exp(-{\bar \lambda}|\tau|)/2\bar\lambda$ as stated
above.  In Fourier space, Eq. (1) amounts to $X_\omega=\eta_\omega
(1+J(\omega))/(i\omega+{\bar \lambda})$, which allows us to relate the Fourier
transform of $f$ to that of $J$: $f(\omega)=|1+J(\omega)|^2/(\omega^2+{\bar
\lambda}^2)$, with the correct large $\omega$ behavior, since $J(\omega)\to 0$
in this limit.

We now proceed to a variational and perturbative calculation of $\bar
\theta$, which will be tested by simulating Eq. (1), before applying
these results to the spin flip problem.

$P(\beta)$ can be written as the ratio of two path integrals, the first one
${\cal Z}_1$ over, say, positive trajectories $X(\tau)$, the second one
${\cal Z}_0$ over unrestricted trajectories:
\begin{equation}
P(\beta)=\frac{2\int_{X>0}{\cal D}X(\tau)\exp [-{\cal S}] }
{\int {\cal D} X(\tau) \exp [-{\cal S}]}=\frac{{\cal Z}_1}{{\cal Z}_0}
\label{two}
\end{equation}
where ${\cal S}={{1}\over {2}} \int_0^{\beta}
\int_0^{\beta} X({\tau}_1) G({\tau}_1-{\tau}_2) X({\tau}_2)\,d\tau_1
d\tau_2$, and $G(\tau_1-\tau_2)$ is the inverse matrix of
$f(\tau_1-\tau_2)$. ${\bar \theta}$ is then calculated from
$P(\beta)$ by taking the limit, ${\bar \theta}=-\lim_{\beta \to \infty}
{{\beta}^{-1}} \ln P(\beta)$. We impose periodic boundary conditions,
$X(0)=X(\beta)$ for the paths, which should not affect the value of ${\bar
\theta}$ in the limit of large $\beta$. We notice that the Gaussian weight
in Eq. (1) then becomes
${\cal S}=\frac{1}{2\beta}\sum_n G(\omega_n)|X(\omega_n)|^2$, where
$G(\omega_n)=1/f(\omega_n)$ and
${\omega}_n={2\pi n/{\beta}}$ are Matsubara frequencies. First consider a
Markov process for which $G(\omega)=
\omega^2+{\bar \lambda}^2$. We recognize the action in imaginary time
($\beta$ is then the inverse temperature) of an
harmonic oscillator of frequency ${\bar \lambda}$, ${\cal S}=\int_0^\beta{\cal
L}(X(\tau))\,d\tau$, with ${\cal L}(X)=\frac 12[(\frac {dX}{d\tau})^2
+{{\bar \lambda}^2} {X^2}]$. Thus, $P(\beta)$ is the ratio between
the partition function of an oscillator with a infinite wall at $X=0$ and
that of the same oscillator without the wall. For large $\beta$, it goes as
$\exp(-\beta(E_1-E_0))$, where $E_1$ ($E_0$) is the ground state energy
of the oscillator with (without) a wall. Thus, $E_0={\bar \lambda}/2$, and
$E_1=3{\bar \lambda}/2$, as the eigenstates of the problem with a hard wall at
the origin are the odd states of the unrestricted oscillator. This gives
the Markovian result ${\bar \theta}={\bar \lambda}$.

For non-Markovian processes, unfortunately, ${\cal S}$ is no longer a classical
action with an associated quantum problem. The denominator ${\cal Z}_0$ can,
however, still be computed exactly and we find after taking
$\beta\to\infty$,
\begin{equation}
E_0= {1\over {2\pi}}\int_0^{\infty}
\ln \left({G(\omega)}\over {\omega^2}\right)\,d\omega
\label{three}
\end{equation}
As a check, one can verify that for an oscillator, for which
$G(\omega)={\omega}^2+{{\bar \lambda}}^2$, one recovers $E_0=
{\bar \lambda}/2$.
The most difficult part is to evaluate the ``ground state energy''
$E_1=-\lim_{\beta \to \infty}\ln {\cal Z}_1/\beta$ of the problem with a
wall at the origin. One way of computing ${\cal
Z}_1$ will be to perturb around a classical action, for instance, that of
harmonic oscillator of frequency $\omega_0$ (or that of a particle in a box
as we also did in \cite{SM1}). We adopt a variational method, by choosing a
trial inverse correlator $G_0(\omega)=\omega^2+\omega_0^2$, corresponding to
that of an oscillator (with a hard wall at the origin) whose  frequency
$\omega_0$ is going to be our variational parameter. We have the
general variational inequality, $E_1\leq 3\omega_0/2 + \lim_{\beta \to
\infty}{1\over {\beta}}{\langle {\cal S}-{\cal S}_{0}\rangle}_{w}$, where
the average is performed using the action of the hard wall oscillator.
The second term of the inequality requires evaluating the propagator
$\langle |X(\omega_n)|^2\rangle_{w}$ of the hard wall oscillator, leading
finally to,
\begin{eqnarray}
&E_{1}^{(2)}=\omega_0 \left[\,{3\over {2}}+{2\over {\pi}}\left ({G(0)\over
{\omega_0^2}}-1\right) \right.\nonumber\\
&+{2\over {\pi}} \left.
\int_0^{\infty}dx \, \left({G(x\omega_0)\over {\omega_0^2}}-x^2-1\right)
\, \sum_{n=1}^{\infty}{n c_n\over{x^2+4n^2}}\right]
\label{four}
\end{eqnarray}
where the numbers $c_n$ can be evaluated using properties of Hermite
polynomials and are given by
\begin{equation}
c_n={4\over {\pi 2^{2n}(2n+1)!}}\,\left [{(2n)!\over {n!(2n-1)}}\right ]^2.
\label{five}
\end{equation}
One can then differentiate Eq. (4) with respect to $\omega_0$ to obtain an
equation for $\omega_0$, which minimizes $E_{1}^{(2)}$. This equation can
then be easily solved numerically. In principle, the value of $E_1$ can be
improved by summing higher terms of the
cumulant expansion around the trial action \cite{SM1}. The superscript
$^{(2)}$ denotes that we have kept only the first two terms of this
expansion.
To perform a systematic order by order cumulant expansion, one should also
keep only the first two terms in $E_0$ (even though $E_0$ can be evaluated
exactly to all orders from Eq. (3)). This gives
\begin{equation}
E_{0}^{(2)}=\omega_0 \left[\frac 12+{1\over {2\pi}} \int_0^{\infty}dx \,
\left (\frac{G(x\omega_0)}{\omega_0^2(x^2+1)}-1\right)\right].
\label{six}
\end{equation}
We can then define ${\bar \theta}^{(2)}$ as ${\bar \theta}^{(2)}=\min_{\omega_0}
(E_1^{(2)}- E_0^{(2)})$, remembering that ${\bar \theta}_{v}=\min_{\omega_0}
E_1^{(2)}-E_0$, is an exact (presumably bad) bound of ${\bar \theta}$.

When $J(\omega)$ is small (see Eq. (1) and below), and  using Eq. (4-6),
one can perform a straightforward perturbative calculation around the
Markov process $J=0$ to first order in $K(\omega) = J(\omega) + J(-\omega)$,
\begin{equation}
\bar\theta={\bar \lambda}\left[1-\frac 2\pi K(0)-{1\over {2\pi}}
\int_0^{\infty}K(x{\bar \lambda})V(x)\,dx\right],
\label{seven}
\end{equation}
where $V(x)=4(x^2+1)S(x)-1$, and $S(x)$ is the same series that appears in Eq.
(4). In fact, an infinite number of terms of the perturbation theory
can be resummed by using a novel technique (for details, see \cite{SM1}),
leading to,
\begin{eqnarray}
\bar\theta=\frac 4\pi \sqrt{G(0)}+{1\over {2\pi}}
\int_0^{\infty}W(G(\omega) / \omega^2)\,d\omega,\\
W(x)=\sum_{n=1}^\infty\frac{c_n}{n}\ln(1+4n^2(x-1))-\ln(x)
\label{eight}
\end{eqnarray}
This expression is valid provided $G(\omega)/\omega^2\geq 1$, which is always
the case for (and close to) a Markovian process. We have tested
Eq. (8), by comparing its prediction to the direct simulation of Eq. (1),
with $f(\tau)=\varepsilon \exp(-\tau)/2+(1-\varepsilon)\exp(-2\tau)/4$,
interpolating between two Markovian processes with ${\bar \lambda}=2$ and
${\bar \lambda}=1$. Note that $\varepsilon$ must be positive to ensure $f>0$,
and that for $\varepsilon>4/3$, one can find $\omega$ such that
$G(\omega)/\omega^2< 1$, and Eq. (8) is not reliable any more. The results
for some representative values of $\varepsilon\ne 0,1$ (for which
$\bar\theta=2$ and $\bar\theta=1$) are displayed in Tab. 1, showing a very
good agreement between Eq. (8) and numerical simulations.

We now turn to the
$T=0$ dynamics of the Ising model starting from a random
(high temperature) initial configuration. We would like to show that
calculating the fraction of unflipped spins up to time $t$ in the Ising model
reduces to calculating the survival probability of a Gaussian process
in the framework of Gaussian closure approximation (GCA), introduced
by Mazenko\cite{M1,LM}. But before we make this connection, a few facts
about the $T=0$ dynamics of the Ising model would be relevant.

Following a quench to $T=0$, interpenetrating domains of $\pm 1$ phases grow
with time. A scaling theory has been developed to characterize the morphology
of the growing domains \cite{AB}. According to this theory, at late times,
the system is characterized solely by the linear length of a growing domain
$L(t)$. For the Ising model, $L(t)\sim t^{1/2}$ in all dimensions (however in
$d=3$ cubic lattice at $T=0$, it seems that $L(t)\sim t^{1/3}$ due to lattice
effects \cite{SS,SM1}, a fact which was underestimated in \cite{sta}). Another
prediction of the scaling theory relevant for us is that the on-site
autocorrelation satisfies $\langle S(t)S(t')\rangle= F(L(t)/L(t'))$ for $t\geq
t'\gg 0$, where $F(x)\sim x^{-\lambda}$ for large $x$ \cite{FH,AB}. The
exponent $\lambda$ is exactly $1$ in $d=1$, is close to $1.25$ in $d=2$ (both
from simulations \cite{FH} and direct experiment \cite{NM}) and close to 1.67
in $d=3$ \cite{AB}. GCA has been particularly successful in calculating this
exponent as it predicts $\lambda_{1d}=1$, $\lambda_{2d}=1.289$ and
$\lambda_{3d}=1.673$, this approach becoming asymptotically exact for large
$d$. Recently, we have extended GCA to the $q$-state Potts model and
calculated the $q$-dependent $\lambda$ \cite{SM}.

Without entering into the details of GCA, we simply mention that this method
assumes that the spin at position $\bf x$ and time $t$ is essentially the sign
of a continuous $Gaussian$ stochastic variable $m(\bf x,t)$, which is
physically interpreted as the distance to the nearest interface to the point
$\bf x$ \cite{M1}. We thus see that in the framework of GCA, \it  the
probability that a spin does not change sign up to time $t$ is equal to the
probability that the associated Gaussian process $m(t)$ does not cross the
origin $m=0$ \rm (from now we forget the label $\bf x$ as this probability
does not depend on the considered site). The process $m(t)$ at a given site is
Gaussian whose correlator $\langle m(t)m(t') \rangle$ can be calculated using
the GCA scheme. However, $m(t)$ is not stationary since its correlation
function depends explicitly on both $t$ and $t'$, and we  need to use the
following trick before using the results of Eq. (4-9). We define the variable
$X(t)= m(t)/\sqrt{\langle m^2(t)\rangle}$ which is also Gaussian. Its
correlator turns out to satisfy $\langle X(t)X(t')\rangle = g(L(t)/L(t'))$,
where once again $g(x)\sim x^{-\lambda}$ for large $x$. If we now set
$\tau=\ln(L(t))$, we are left to study the survival probability of a
$stationary$ Gaussian process, with correlator $\langle
X(\tau)X(\tau')\rangle= f(\tau-\tau')$, with $f(\tau)=g(\exp|\tau|)$. This
completes the relation to the general problem studied in the first part of
this Letter. For the Gaussian process, the survival probability
$P(\tau)\sim \exp(-{\bar \theta}\tau)\sim L(t)^{-{\bar \theta}}$. Thus we
get $\theta = {\bar \theta}$ in the framework of GCA. Note that for an
ordinary Brownian walker $g(x)=x^{-1/2}$, or
$f(\tau)=\exp(-|\tau|/2)$, which leads to the well known result $P(t)\sim
t^{-1/2}$.

By means of GCA (see \cite{SM1} as these calculations are not the subject of
this Letter), we have computed the functions $f$ for $d=1,2,3$. For instance,
in $d=1$, we got the (exact) result, $f(\tau)= \sqrt{2/ {(1+\exp
(|2\tau|))}}$. We see that $f$ is not a pure exponential and is thus
non-Markovian. Knowing $f(\tau)$, we have computed the variational
$\bar\theta^{(2)}$, and the associated frequency $\omega_0$. In Tab. 2, these
results are compared to $\bar\theta_{NS}$, extracted from numerical
simulations of Eq. (1) (in Fourier space, then using inverse FFT), with the
associated $f$. We also performed Ising simulations in $d=2$ ($800\times 800$
lattice, 30 samples) confirming the results of \cite{D1,sta,BD2}), and in $d=3$
($100\times 100\times 100$ lattice, 60 samples). Note that the scaling of
$P(t)$ is found to be much better as a function of $L(t)$ than as function of
$t$ \cite{SM,BD2}.

We note that the variational results are in good agreement with the
simulations of Eq. (1), and with the Ising results in $d=1$. However, the
agreement in $d=2,3$ is not as good. To understand this, we have to remember
that there are two possible sources of errors between the simulations of Eq.
(1) and the Ising simulations: $(i)$ the real variable $X(t)\sim m(t)$ is not
truly Gaussian and $(ii)$ due to the use of GCA, the correlator $f$ is not
exact (except in $d=1$, by accident). To illustrate the importance of the
error due to $(i)$, notice that although $f$ is exact in $d=1$, the Ising exact
result $\theta=3/4$ is bigger than the value 0.70 obtained by simulating
Eq. (1), which is perfectly reproduced by our theory. Note that, we got the
exact bound for $\bar\theta$, $\bar\theta_v=0.736$ in $d=1$,
whereas $\bar\theta_v$ is a bad bound for $d>1$.
In $d=2,3$, the main cause of error is presumably $(ii)$, as $f$ given by
GCA is found to decrease faster than in the Ising simulations for $\tau <2$
\cite{SM1}. To check this, we computed $\bar\theta$ from Eq. (1) and
$\bar\theta^{(2)}$ using for $f$ a fit of the correlator obtained in the
Ising simulations, leading to
a clear improvement (see Tab. 2). In $d=3$, the numerical $f(\tau)$ can be
computed only up to a quite short time $(\tau<1.5)$, and $\theta$ is quite
dependent on the way $f(\tau)$ is extended to large $\tau$. Indeed, for such
short times, the known asymptotics $f(\tau)\sim\exp(-\lambda\tau)$ is not
reached yet. Finally, one can easily understand why the theoretical estimates
in $d=2,3$ seem less accurate than in $d=1$ or in Tab. 1. It is possible to
show \cite{SM1} that for the Ising model $f$ becomes {\it less and less
Markovian} \rm as $d$ increases. An illustration of this is the fact that,
as $d$ increases, the relative difference between $\theta$ and $\lambda$
strongly increases. In addition, the region for which $G(\omega)/\omega^2<1$
becomes wider as $d$ increases, which is also a sign of strong
``non-Markovianity'', and which prevented us from using Eq. (8) to evaluate
$\theta$. However, for large $d$, an alternate approach has been
derived \cite{MSBC,SM1} which gives $\theta$ in excellent agreement with
numerical simulations.

In summary, we have implemented variational  and perturbative approaches
to compute the survival probability of a Gaussian non-Markovian process, close
enough to the Markovian limit associated to a quantum problem. These results
were then used to calculate the fraction of unflipped spins in the zero
temperature Monte-Carlo dynamics of the Ising model within the framework of
GCA. Further details of these calculations, the application to the calculation
of the $q$-dependent $\theta$ for the $q$-state Potts model \cite{SM,D1,BD2},
numerics and other approaches will be presented elsewhere \cite{SM1}. Finally
the related question of the probability that the $total$ magnetization never
flips after a quench at $T=T_c$ \cite{MBCS} or at $T=0$ \cite{SM1} is also the
subject of further studies.

\acknowledgements
We thank J. Bellissard, A.J. Bray, S. Cueille, D. Dhar, B. Derrida, C.
Godr\`eche, V. Hakim, L. Miclo, N. Read, S. Sachdev, R. Shankar  and R. Zeitak
for useful discussions. S.M. also thanks I. Gruzberg, K. Damle and T. Senthil
for illuminating remarks and the hospitality at the Universit\'e Paul Sabatier
(Toulouse) where this work was initiated. The $d=3$ value for $\theta$ has
been also computed by S. Cueille on $200^3$ lattices. S.M.'s research was
funded by NSF grant \# DMR-92-24290 and CNRS.

\table{Tab. 1: $\bar\theta_{NS}$ obtained from direct numerical simulations
(NS) of Eq. (1)
($\tau_{max}\sim 1000$, $\Delta \tau\sim 10^{-3}$, 100 samples, uncertainties
are $\Delta\theta=\pm 0.015$) is compared to the result of Eq. (8) as a
function of $\varepsilon$, for $f(\tau)=\varepsilon \exp(-\tau)/2 +
(1-\varepsilon) \exp(-2\tau)/4$. We have reported the point $\varepsilon=4/3$
(beyond which Eq. (8) must be regularized (see text)), which shows that
$\theta$ can be $less$ than the smallest inverse relaxation time (here 1).
Note the remarkable accuracy of Eq. (8), up to its validity limit.}

\table{Tab. 2: the four columns contain respectively, 1) $\theta_{Ising}$
obtained in our Ising model numerical simulations for $d=2,3$; 2)
$\bar\theta_{NS}$ obtained by simulating Eq. (1) ($\tau_{max}\sim 1000$,
$\Delta \tau\sim 10^{-3}$, 500 samples); 3) $\bar\theta^{(2)}$ from Eq.
(7-8); 4) the optimal frequency $\omega_0=\bar\theta^{(1)}$ which can be
interpreted  as the first term in the cumulant expansion. For $d=2,3$, we also
give the same results using a fit of $f$ taken from the Ising simulations,
instead of GCA. In $d=3$, $\theta$ depends on the nature of the fit (see
text) and we only give typically obtained values (estimated uncertainty of
order 0.05).}

\newpage
\centerline{\bf TABLE 1}
\vskip 2cm
\begin{tabular}{|l|l|l|l|l|l|}

$\varepsilon$ & {$0.10$}
& {$0.25$} & {$0.5$} & {$0.75$} & {$1.33$}  \\ \hline
$ {\bar\theta}_{NS}$ & $1.72$ & $1.47$ & $1.22$ & $1.09$ & $0.91$
\\ \hline
${\bar\theta}_{Theory}$ & $1.74$ & $1.48$ & $1.24$ & $1.09$ & $0.91$
\\  \hline
\end{tabular}

\vskip 5cm
\centerline{\bf TABLE 2}
\vskip 2cm
\begin{tabular}{|l|l|l|l|l|}

{} & ${\theta}_{Ising}$ & ${\bar\theta}_{NS}$
& ${\bar\theta}^{(2)}$ & ${\bar\theta}^{(1)}$ \\ \hline

$d=1$ & $3/4$ & $0.70\pm 0.01$ & $0.70$ & $0.61$ \\ \hline

$d=2\,$ (GCA) & $0.45\pm 0.01$ & $0.66\pm 0.01$ & $0.58$ & $0.43$ \\
$d=2\,$ (fit of $f$) & -- & $0.49\pm 0.02$ & $0.38$ & $0.27$ \\ \hline
$d=3\,$ (GCA) & $0.52\pm 0.01$ & $0.70\pm 0.02$ & $0.47$ & $0.30$ \\
$d=3\,$ (fit of $f$) & -- & $\sim 0.55$ & $\sim 0.4$ & $\sim 0.2$ \\ \hline
\end{tabular}

\end{document}